\documentclass[10pt,pre]{revtex4}

\RequirePackage{andreas}
\RequirePackage{graphicx}

\begin{document}

\author{Andreas Kabel}
\thanks{Work supported by Department of Energy contract
DE--AC03--76SF00515.}
\affiliation{Stanford Linear Accelerator Center,\\
  2575 Sand Hill Road, Menlo Park, CA~94025\\
E-Mail: andreas.kabel@slac.stanford.edu}

\title{Collective quantities of a fermionic particle beam 
       \\in a circular accelerator near zero temperature}

\begin{abstract}
  In the usual parameter regime of accelerator physics, particle
  ensembles can be treated as classical. If we approach a regime where
  $\epsilon_x\epsilon_y\epsilon_s \approx
  N_{\textrm{particles}}\lambda_{\textrm{Compton}}^3$, however, the granular structure
  of quantum-mechanical phase space becomes a concern. In particular,
  we have to consider the Pauli exclusion principle, which will limit
  the minimum achievable emittance for a beam of fermions.  We
  calculate these lowest emittances for the cases of bunched and
  coasting beams at zero temperature and their first-order change rate
  at finite temperature.  The self-field of the particle beam will,
  analogous to a space-charge tune depression, lead to a decrease of
  the quantum-mechanical state density. We calculate the tunes and the
  increased emittance for this case.
\end{abstract}

\maketitle

\newcommand{\energy}{\epsilon}
\newcommand{\efermi}{\energy_F}
\newcommand{\emittance}{\varepsilon}
\newcommand{\dhalf}{\frac{\tilde d}2}

\section{Introduction}

When a relativistic particle travels in a focusing dominated beamline,
it can be viewed, in its rest frame, as a particle in a harmonic
oscillator potential. Emission of dipole radiation will lead to
transverse energy loss; quantum-mechanically, the particle will drop
down one energy level by emission of a photon.  This mechanism will,
in the absence of competing heating mechanisms, ultimately lead to a
cooling down of the particle's transverse action to its
quantum-mechanical limit of $J_{\textrm{min}}=\frac h2$. This result
has been established in \cite{huang:1998}, however, only
single-particle dynamics is considered there.

In the multi-particle case, some care has to be taken if the particles
in question are fermions, as Pauli's exclusion principle has to be
considered when constructing the ground state; it has been claimed
elsewhere\cite{Mikhailichenko:1999dm} that this will limit the
admissible minimum emittance to values not too far out of the reach of
today's technologies.

In this paper, the emittances for a relativistic fermion bunch and a
coasting beam are calculated for temperatures close to zero. We will
show that the quantum-mechanical limit for the transverse emittances
for typical configurations are vastly smaller than values reached
today.

\section{Calculating collective quantities: the Wigner transformation}
\label{section:wigner}

In this paper, we want to calculate the emittance of a multi-particle
system. The classical emittance is defined in terms of classical
expectation values, \abk ie particle averages, as
$\emittance^2=\expect{p^2}_{\textrm{Cl}}\expect{q^2}_{\textrm{Cl}}-\expect{pq}^2_{\textrm{Cl}}$.
A quantum version of this and similar quantities can be obtained by
interpreting the Wigner transform of the state of the quantum system
in question as a classical probability distribution.

Consider an $N$-particle system in a pure state $\ket{\Psi} =
P\ket{n_1}\cdots \ket{n_n}$, where $P$ is the (anti)symmetrizing
operator, the particles in question being (fermions) bosons. Now, we
can create a pseudo-classical probability distribution
$\rho(q_1,p_1;\ldots;q_N,p_N)$ by using the Wigner transform of the product
wavefunction.  It is easy to see that all expectation values of this
phase-space distribution are the same as the ones of the Wigner
transform of a one-particle statistical matrix with equal weights in
all states. We can thus conclude that the  appropriate quantum version
of the classical expectation values above is obtained by replacing the classical
average $\expect{f(p,q)}$ by
$\expect{\expect{f(p_{op},q_{op})}_{\textrm{Qm}}}_{\ket{n_i}}$, \abk ie, the
quantum-mechanical expectation value for the corresponding operator
function, averaged over the occupied states.

\section{Dynamics}

Consider an ultra-relativistic particle beam in a circular
accelerator.  Neglecting higher-order effects, the Hamiltonian can be
written as a quadratic form in the usual phase-space co{\"o}rdinates
$x,x',y,y',\sigma,\delta$.

However, this Hamiltonian is not appropriate for quantization, as
energy and time have switched roles. Thus, we use the Hamiltonian of
the system in the beam's frame of reference, which can be obtained by
a series of canonical transformation from the lab
frame\cite{Wei:1993}:
\begin{equation}\label{eq:hamiltonian}
  H = \frac{p_x^2}2 + 
      \frac{p_y^2}2 + 
      \frac{p_z^2}2 -
      \frac{\gamma^2\beta xp_z}R + 
      \beta^2\gamma^2\left(\kappa_x-\frac{\gamma^2}{R^2}\right)\frac{x^2}2 + 
      \beta^2\gamma^2\kappa_y\frac{y^2}2  + \Phi_{RF}(z)\satzz,
\end{equation}
where $\gamma$ is the relativistic factor, $\kappa_{x,y}$ the
(possibly local) focusing strengths (in the case of magnetic
quadrupoles, one has $\kappa_x = -\kappa_y$) and $\Phi(z)$ the
external electric potential. The directions $x,y,z$ are radial,
transversal, and tangential, respectively.  Note that we use units
with $\hbar=c=k_B=m_0$ throughout, so all quantities are expressed in
powers of the Compton length of the particles in question.

The longitudinal part of the Hamiltonian depends on the physical
setup. The particles might either be confined by the nearly harmonic
potential of the RF bucket, or we have case of a coasting beam, where
the only constraints imposed on the longitudinal motion are the ones
due to the periodicity of the problem. In the sequel, we will consider
both cases.

\section{Anisotropic Oscillator}

Let us assume that the longitudinal motion is determined by the
presence of an RF bucket. We can approximate the potential $\Phi$ by
expanding it to 2nd order in $z$. For reasons of simplicity, we only
take into account the $\Order(z^2)$ term, \abk ie, we assume that the
particle is on the orbit and is not losing energy.

For a bunched beam with dimensions
$\sigma_z,\sigma_\delta$, we have, in the orbit
frame, $\omega_l\gamma\sigma_l = \sigma_{\delta}$.


The longitudinal and radial parts of the hamiltonian (\ref{eq:hamiltonian}) have the
form
\begin{equation}\label{eq:hamiltonianII}
  H = \sum_{i=1}^2{\frac{p_i^2}2 + \frac{\omega_i^2q_i^2}2} - \mu q_1 p_2
 \satzz.
\end{equation}
%
Its eigenfrequencies are determined by the equation
\begin{equation}{\label{eq:eigenfreq}}
  \lambda^4+\lambda^2(\omega_1^2+\omega_2^2)+\omega_2^2(\mu^2-\omega_1^2)=0
  \satzz,
\end{equation}
which leads to stable motion for
\begin{equation*}
  (\omega_1^2+\omega_2^2)>4\omega_2^2(\mu^2-\omega_1^2)>0\satzz.
\end{equation*}

The first condition will always be fulfilled for realistic machines.
The second one corresponds to the machine being below or above
transition: if the second factor changes sign, the eigenfrequency can
be mad real again by flipping the sign of
$\omega_z^2\propto\Phi_{RF}''$.  However, the absolute sign of both
the kinetic and the potential term will change, leading to the (for
purposes of constructing the quantum-mechanical ground state)
pathological case of a hamiltonian not limited from below. In the
sequel, we will assume the machine is below transition.

Thus, expanding (\ref{eq:hamiltonian}) to first non-trivial order in
the canonical co{\"o}rdinates and applying the canonical
transformation removing the mixed term in (\ref{eq:hamiltonianII}), we
obtain the Hamiltonian of a 3-dimensional harmonic oscillator with
corrected frequencies given by (\ref{eq:eigenfreq}); the ground state
is characterized by the occupation numbers $n_n^d\in\{0,1\}$ where
$\sum n_d^i=N$ of the oscillator levels.  For sake of generality, we
consider the case of $d$ dimensions.  The ground state for a given
particle number can be constructed by successively filling states with
the lowest energy (we disregard spin here, which can be easily
reintroduced by replacing $N\to2N$ in the final formulae).

In $\frac E{\efermi}$-space, the Fermi sea is just a unit $d$-simplex,
in $n_i$-space, a $d$-simplex with axes of length
$\frac{\omega_1}{\efermi},\ldots\frac{\omega_d}{\efermi}$. Thus, the
particle number for a ground state filled up to the Fermi energy
$\efermi$, where we have disregarded the zero-mode energy
$\frac12\sum_i\omega_i$ of the oscillator,
\begin{equation*}
  N = \frac{\efermi^d}{\Omega d!}\satzz,
\end{equation*}
the volume of an unit $d$-simplex being $\frac1{d!}$ and
$\Omega={\omega_1\cdots\omega_d}$.

The energy in the $i$th degree of freedom in that case is given by a
sum over the $d$-simplex. Replacing all sums by integrals, we have
\begin{equation*}
  E_i = \frac{\efermi^d}{\Omega}\int_{0}^{1}
      \int_{0}^{1-q_1}\cdots
      \int_{0}^{1-\sum_{k=1}^{d-1}q_k}\efermi q_i\td q_1\cdots\td q_d =
      \frac{\efermi^{d+1}}{\Omega(d+1)!}=\frac{N\efermi}{d+1}\satzz.
\end{equation*}

We calculate the emittance by using its statistical definition
$\emittance^2=\expect{p^2}{q^2}-\expect{pq}^2$ and replacing the
classical averages by the ones discussed in \ref{section:wigner}.

For a harmonic oscillator, we have $\expect{p^2}_{\textrm{Qm}}\expect{q^2}_{\textrm{Qm}}=n$ and
$\expect{pq}_{\textrm{Qm}}=0$, so:
\begin{equation}\label{emitt:quadr}
   \emittance_i = \expect{n_i}_{\ket{n_i}} = 
   \frac{E_i}{N\omega_i}= \frac{\efermi}{\omega_i(d+1)} = 
   \frac{\Omega^{\frac1d}}{\omega_i(d+1)}\sqrt[d]{Nd!}
   \satzz.
\end{equation}
and the total phase-space volume
\begin{equation*}
    \begin{aligned}
      \emittance^{(d)} &= \prod_{i=1}^d\emittance_i = \frac{Nd!}{(d+1)^d} \\ 
      \emittance^{(3)} &= \frac3{32}N
    \end{aligned}
\end{equation*}

Thus, the projected emittances scale as $N^{\frac1d}$, as one would
na{\"\i}vely assume.  Furthermore, due to the occurrence of the
geometric mean of the frequencies in (\ref{emitt:quadr}), the
projected emittance in one dimension can be lowered by shallowing the
potential in the other dimensions.

Note that a similar approach has been chosen elsewhere;
[\cite{Mikhailichenko:1999dm}] gives an estimate for
$\emittance_{min}$ from a similar reasoning, but ends up (due to a
miscounting of the states) with a scaling different from our result.

\section{Mixed Case: Longitudinally Free Particles}

So far, we have assumed an anisotropic oscillator. But given the case
of a particle moving freely longitudinally, the energy content of that
degree of freedom will be given by the square of the (angular)
momentum.  (We might consider the boundary conditions imposed by a
periodic box instead of a circular arrangement.) We treat the general
case, \abk ie, a hamiltonian
\begin{equation*}
  H = \sum_{\tilde i=1}^{\tilde d} 
      \tilde\omega_{\tilde i}^2 \tilde n_{\tilde i}^2
   + \sum_{i=1}^d \omega_i n_i\satzz.
\end{equation*}

The emittance in each of $\tilde d$ new degrees of freedom is given by
$\expect{p^2}_{\textrm{Qm}}=\frac{n^2\pi^2}{D^2}=2E_n$ and
$\expect{q^2}_{\textrm{Qm}}=\frac{D^2}{12}$, where $D$ is the length of the
enclosing potential, and $\expect{pq}_{\textrm{Qm}}=0$, so
\begin{equation*}
  {\tilde\emittance}_{\tilde i} = 
  D\sqrt{\frac{\expect{\tilde E_{\tilde i}}_{\ket n}}6}=
  \sqrt{\expect{{\tilde E}_{\tilde i}}_{\ket n}}
  \frac{\pi}{\sqrt{12}{\tilde \omega}_{\tilde i}}
\satzz.
\end{equation*}
Rescaling the integration range to a unit sphere and unit simplex, we get
\begin{equation}\label{eq:efermimixed}
  N = \frac{\efermi^{\dhalf+d}}
           {\tilde\Omega\Omega}
    \int_{\tilde d\textrm{-sphere}} 
    \int_{d\textrm{-simplex}}^{1-\tilde q^2}\td q\td{\tilde q}
    =  \frac{\pi^{\dhalf}\efermi^{\dhalf+d}}
         {\tilde\Omega\Omega\Gamma\left(\frac{2d+\tilde d+2}2\right)}
\satzz,
\end{equation}
%

We can readily write down the averaged values of the energy in
the different degrees of freedom:
\begin{equation*}
  \frac{\expect{{\tilde E}_{\tilde i}}}{\efermi}=\expect{\tilde q^2}=\frac
  {\int_0^1 {\tilde q}^2(1-\tilde q^2)^{\frac{d-1}2+d} \td \tilde q}
  {\int_0^1 (1-\tilde q^2)^{\frac{d-1}2+d} \td \tilde q}
  = \frac1{2d+\tilde d+2}
\end{equation*}
and
\begin{equation}\label{eq:transverseenergy}
  \frac{\expect{E_{i}}}{\efermi}=\expect{q} = 
  \frac
  {\int_0^1q(1-q)^{\dhalf+d-1}\td q}
  {\int_0^1(1-q)^{\dhalf+d-1}\td q}
  =\frac{2}{2d+\tilde d+2}
\end{equation}
so
\begin{equation}\label{eq:emittancescaling}
  \emittance_i = 
  \frac{\efermi}{\omega_i}
  \cdot
  \frac{2}{\tilde d+2d+2}\propto
  N^{\frac2{\tilde d+2d}}
\end{equation}
and
\begin{equation*}
  {\tilde\emittance}_{\tilde i} = \sqrt{\frac{\efermi}
    {{\tilde \omega}_{\tilde i}}}\cdot\frac\pi{\sqrt{12}(\tilde d+2d+2)}\propto
  N^{\frac1{\tilde d+2d}}
\end{equation*}
and the product emittance is 
\begin{equation*}
  \emittance^{(d,\tilde d)} = 
  \frac
  {\pi^{\tilde d}2^d\efermi^{d+\frac d2}}
  {12^{\dhalf}(\tilde d+2d+2)^{d+\dhalf}\Omega\tilde\Omega}=
  \left(
    \frac{\pi}{24}
  \right)^{\dhalf}
  \cdot
  \frac
  {\Gamma\left(\dhalf+d+1\right)}
  {\left(\dhalf+d+1\right)^{\dhalf+d}}N\satzz.
\end{equation*}
Of course, by putting $\tilde d=0$, we regain the formulae for the bunched-beam case.

Putting in a real coasting-beam ring, we can express the orbit-frame
frequencies by the tune\cite{Venturini:Oscillator}:
$\omega_x\approx\omega_y=\frac{\beta\nu_y}{\gamma L}$, where $L$ is
the length of the ring. The longitudinal momentum is quantized in
units of $\frac{\pi}{\gamma L}$, so $\omega_l =
\frac{\pi}{\sqrt{2}\gamma L}$ and
\begin{equation*}
  \emittance_{x} = \expect q \frac{\efermi}{\omega_{x}} = 
  \frac17\sqrt[5]{\frac{225\pi N^2}{256\gamma L\nu}}
  \satzz.
\end{equation*}

This means the transverse degrees of freedom begin to exhibit a non-minimal
emittance for $N>N_0$, where
\begin{equation*}
  N_0 \approx 78.0\sqrt{\frac{\nu\gamma L}{\lambda_{\textrm{Compton}}}}
\end{equation*}
in usual units, and the transverse emittances will grow $\propto N^{\frac 25}$
for higher particle numbers.

For parameters one could consider for a real focusing-dominated,
below-transition ring ($\gamma=10,\nu=100,L=2\pi{\textrm m}$), we get
$N_0\approx1.0\cdot10^{10}$ particles; \abk ie, the beam will have, for
realistic particle numbers, an emittance close to its minimum value.
($\lambda_{\textrm{Compton}}$ in customary units.)



\newcommand{\weff}{\omega_{\textrm{eff}}}
\newcommand{\wext}{\omega_{\textrm{ext}}}
\newcommand{\wSC}{\omega_{\textrm{SC}}}
\newcommand{\wint}{\omega_{\textrm{int}}}

\section{Self-consistent tune shift for the Fermi condensate}

In our construction, we tacitly assume that the particle-particle
interaction does not modify the particle content of the ground state.
This corresponds precisely to the notion of a Fermi liquid (in our
case, a highly anisotropic one), in which the free particle spectrum
smoothly deforms into the quasi-particle spectrum of equal particle
content when the interaction is switched on adiabatically. This
na{\"\i}ve assumption of the existence of a Fermi surface may break
down if we take into account particle-particle interactions.

For the case of the system being below transition, we can make
the following semi-quantitative argument for the existence of
a Fermi liquid:
In a ``mean-field calculation'', we estimate the effective transverse
focusing strength $\weff$ to be the sum of the
external focusing and a space-charge tune depression due to a circular beam
of radius $\sqrt{\expect{x^2}}$:
\begin{equation*}
\weff^2=\wext^2-\wSC^2  = \wext^2 - \frac{Ne^2}{2\gamma L\expect{x}^2}.
\end{equation*}

In the follwing, we consider a coasting beam. We have
$\expect{x^2}\weff^2= \expect{E_x}$, where $\expect{E_x}$ is given by
(\ref{eq:transverseenergy}).  We use (\ref{eq:efermimixed}) with
$\omega=\weff$ to eliminate $\efermi$ and obtain a consistency
condition for $\weff$:

\begin{equation}\label{eq:consistency}
  \left(\frac{\weff}{\wext}\right)^2
  \left[
    1 + 
    \left(\frac{\wint}{\wext}\right)^{4/5}
    \left(\frac{\weff}{\wext}\right)^{-4/5}
  \right] = 1\satzz,
\end{equation}
where we have introduced 
\begin{equation}\label{eq:wint}
  {\wint}=\wext\frac{\sqrt[4]{2^37^5
                      r_{\textrm{cl}}^5({\gamma L})N^3}}{\nu\sqrt{15\pi^3\lambdabar_{\textrm{Compton}}^3}}\satzz.
\end{equation}

(\ref{eq:consistency}) has a solution ${\weff}<{\wext}$ for for all $N$.
This corrected frequency has to be substituted in the equations in the
previous section, leading to corrected values for the emittances.

For typical setups, ${\wint}\gg{\wext}$  

(using the ring parameters from above and $N=10^{10}$
particles, we end up with $\wint/\wext\approx 2.14\cdot10^6$ and
$\emittance_{i,{\textrm{corrected}}}\approx 18.5\emittance_i$), 
so
$\frac{\weff}{\wext}\approx\left(\frac{\wint}{\wext}\right)^{-\frac23}$.
The scaling of the transverse emittances is $\propto\weff^{\frac15}$;
with (\ref{eq:wint}), $\weff\propto N^{-\frac12}$, so we end up with a
corrected scaling for the case of large particle numbers:
$\emittance_i \propto N^{\frac12}$ instead of $\emittance_i \propto
N^{\frac25}$ from (\ref{eq:emittancescaling}).
 
This modified scaling law means that the transverse phase space volume
scales as $\emittance_x\emittance_y=\Order(N)$: adding a new particle
to the coasting beam's ground state will lower the transverse
frequencies, thus increasing the state density, so a new site becomes
available in transverse phase space, while the longitudinal phase
space retains its original volume. In particluar, this means that
kinetic and potential energy always will balance out, $\efermi \approx
\energy_{kin} \approx \energy_{pot}$, keeping the system at the
boundary of a ``weakly non-ideal'' and a ``strongly non-ideal'' gas.
Thus, the simple criterion $\energy_{kin}\ll\energy{pot}$ for quantum
Wigner crystallization is not applicable for any external focusing
strength. This property is peculiar to the external oscillator
potential and is not present for other confining potentials.
\begin{figure}
\begin{center}\includegraphics[width=0.8\linewidth]{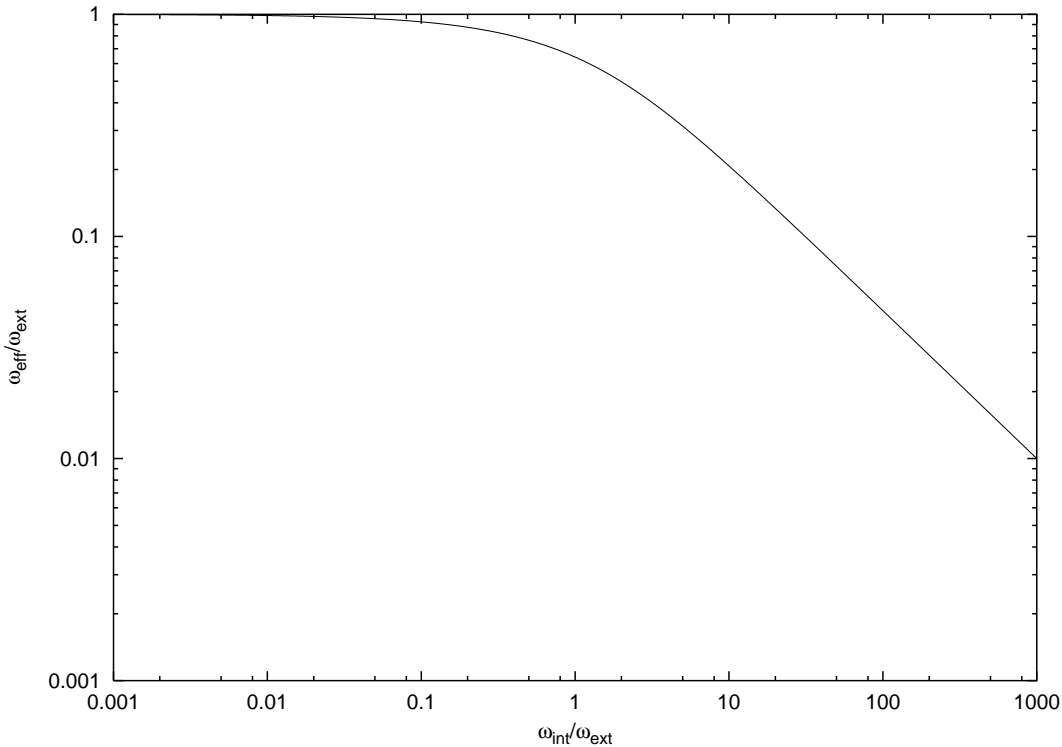}\end{center}
\caption{Effective focusing strength of a coasting beam in its ground state;
 $\omega_{int} = {e^3N}/(\gamma L)$ }\label{fig:effective}
\end{figure}

\section{Finite Temperature}

The above considerations were for the case of zero temperature. To generalize
to finite temperatures, we follow the usual prescription and introduce a 
chemical potential. The quantity
we want to calculate is the logarithm of the partition function of
the grand-canonical ensemble:
\begin{equation*}
\log Z = 
\log \sum_{n_{\vec i}\in\{0,1\}} 
e^
{-\beta 
 \sum_{\vec i}
      n_{\vec i}
      \left(
        \sum_{k=1}^d\omega_k(i_k+\frac12)
        -\mu
      \right)
}    = 
    \sum_{i_k=0}^\infty
    \log\left(
      1+
      e^{
        -\beta
        \left(
          \sum\omega_k(i_k+\frac12)-\mu
        \right)
      }
      \right)
      \satzz.
\end{equation*}

Again, we transform the sum into an integral. With (\ref{eq:efermimixed}),
we integrate over energy
\begin{equation*}
\log Z = 
\frac1{\Omega\tilde\Omega\Gamma\left(d+\dhalf\right)} 
\int_0^\infty 
\log
\left(
  1+e^{-\beta(E-\mu)}
\right)E^{d+\dhalf-1}\td E\satzz.
\end{equation*}
For small temperatures, integrals of this style can be done by
integrating by parts and then be approximated using a Sommerfeld
expansion. We find
\begin{equation*}
  \log Z = -\frac{\beta\pi^{\dhalf}}{\Gamma\left(d+\dhalf+1\right)}\left(\frac{\mu^{d+1}}{d+\dhalf+1} + 
  \left(d+\dhalf\right)\frac{\mu^{d-1}\pi^2}{6\beta^2} + \ldots\right)\satzz.
\end{equation*}
As
\begin{equation*}
  \begin{braced}
    \begin{aligned}
      \emittance_i &= -\frac1{\beta N}
      \pa{\omega_i}\log Z \\
      \tilde\emittance_{\tilde i} &= 
      \sqrt{-\frac{\pi^2}{24\beta N\omega_{\tilde i}}
        \pa{\tilde\omega_{\tilde i}}\log Z} 
\end{aligned}
\end{braced}
\satzz,
\end{equation*}
we can write the temperature-dependent contributions to
the emittances:
\begin{equation}\label{eq:emittancesusc}
\begin{braced}
\begin{aligned}
\frac{\Delta\emittance_i}{\emittance_i} = \frac{T^2\pi^2}{6\efermi^2}
\left(d+\dhalf\right)
\left(d+\dhalf+1\right) 
\\
\frac{\Delta{\tilde\emittance}_{\tilde i}}{{\tilde\emittance}_{\tilde i}} = \frac{T^2\pi^2}{12\efermi^2}\left(d+\dhalf\right)\left(d+\dhalf+1\right)
\end{aligned}
\end{braced}
\end{equation}
where we have used the zero-temperature $\efermi$ as chemical potential, which is correct in this order of $\beta$. Thus, the
emittances grow quadratically in the temperature, with the scale being set
by $\efermi$. For the parameters from above and $N=10^{10}$ particles,
one finds, from (\ref{eq:emittancesusc}), a transverse emittance-doubling temperature of $T=0.03K$.

\section{Acknowledgments}
I wish to thank R.~Ruth and M.~Venturini for useful discussions.

\bibliography{/home/akabel/texlib/bibliography}

\end{document}